\documentclass[galaxies,article,accept,moreauthors,pdftex]{Definitions/mdpi} 

\firstpage{1} 
\pubvolume{9}
\issuenum{4}
\articlenumber{107}
\pubyear{2021}
\copyrightyear{2021}
\externaleditor{Academic Editors: Francesca Loi and Tiziana Venturi}  
\datereceived{6 October 2021} 
\dateaccepted{16 November 2021} 
\datepublished{19 November 2021} 
\hreflink{https://doi.org/\newline10.3390/galaxies9040107}

\newcommand{\farcs}{.\!\!^{\prime\prime}}

\Title{How Are Red and Blue Quasars Different?\newline The Radio~Properties}

\TitleCitation{How Are Red and Blue Quasars Different? The Radio Properties}

\Author{{Victoria A. Fawcett} $^{1,}$*$^{,\dagger}$\orcidA{}, {David M. Alexander} $^{1,\dagger}$\orcidB{}, {David J. Rosario} $^{2,1,\dagger}$\orcidC{} and {Lizelke Klindt} $^{1,\dagger}$\orcidD{}}

\AuthorNames{Victoria Fawcett, David Alexander, David Rosario and Lizelke Klindt}

\AuthorCitation{Fawcett, V.A.; Alexander, D.M.; Rosario, D.J.; Klindt, L.}

\address{%
$^{1}$ \quad Center for Extragalactic Astronomy, Department of Physics, Durham University, {Durham} DH1 3LE, UK; d.m.alexander@durham.ac.uk (D.M.A.); david.rosario@newcastle.ac.uk (D.J.R.); lizelkeklindt@gmail.com~(L.K.) \\
$^{2}$ \quad School of Mathematics, Statistics and Physics, Newcastle University, {Newcastle upon Tyne} NE1 7RU, UK}
\corres{Correspondence: victoria.fawcett@durham.ac.uk}

\firstnote{These authors contributed equally to this work.}

\abstract{
A non-negligible fraction of quasars are red at optical wavelengths, indicating (in the majority of cases) that the accretion disc is obscured by a column of dust which extinguishes the shorter-wavelength blue emission. In this paper, we summarize recent work by our group, where we find fundamental differences in the radio properties of SDSS optically-selected red quasars. We also present new analyses, using a consistent color-selected quasar parent sample matched to four radio surveys (FIRST, VLA Stripe 82, VLA COSMOS 3\,GHz, and LoTSS DR1) across a frequency range 144\,MHz--3\,GHz and four orders of magnitude in radio flux. We show that red quasars have enhanced small-scale radio emission ($\sim$kpc) that peaks around the radio-quiet threshold (defined as the ratio of 1.4\,GHz luminosity to 6~$\upmu$m luminosity) across the four radio samples. Exploring the potential mechanisms behind this enhancement, we rule out star-formation and propose either small-scale synchrotron jets, frustrated jets, or dusty winds interacting with the interstellar medium; the latter two scenarios would provide a more direct connection between opacity (dust; gas) and the production of the radio emission. In our future study, using new multi-band uGMRT data, we aim to robustly distinguish between these scenarios.}


\begin{document}

\section{Introduction}
Quasi-stellar objects (QSOs), also known as quasars, are the most powerful class of Active Galactic Nuclei (AGN). Their extremely high bolometric luminosities (up to $10^{{47}{\text{--}}{48}}$~erg~s$^{-1}$) are now known to be caused by accretion onto a supermassive black hole (SMBH; masses of $10^8$--$10^9$~M$_{\odot}$) near the Eddington limit, which places them as some of the most luminous objects in the Universe. Due to the unobscured view of the SMBH accretion disc, which peaks in the ultra-violet (UV), the~majority of Type~1 QSOs have very blue optical colors. However, there is a small but significant subset with redder optical-infrared colors (referred to as ``red QSOs''). 

Although red QSOs have been well studied in the literature, there are still conflicting views on how red and blue QSOs are related \citep{rich,glik,georg,Urrutia_2009,ban12,glik12,kim18,klindt,fawcett,rosario,rosario_21,calistro}. Some studies suggest they are intrinsically the same objects, with~a red QSO simply representing a blue QSO observed at a higher inclination to the putative dusty torus, while other studies suggest an evolutionary scenario whereby a red QSO is a short-lived phase in QSO evolution. We have recently found that Sloan Digital Sky Survey (SDSS) red QSOs show $\sim$2--3 times higher radio-detection fraction compared to typical blue QSOs, which cannot be explained via a simple orientation model \citep{klindt,fawcett,rosario,rosario_21}. This radio-detection enhancement is driven by red QSOs with compact radio morphologies ($<$5$''$; $<$43~kpc at $z$\,$=$\,1.5) and luminosities placing them around the radio-loud/radio-quiet threshold~\cite{klindt}. We confirmed these initial results with deeper and higher resolution data from the Very Large Array (VLA) Stripe~82, VLA COSMOS 3\,GHz, and LOFAR surveys \citep{fawcett,rosario}. Pushing to lower values of radio-loudness, we found that the radio enhancement decreases to unity at extreme radio-quiet values. Analyzing a sample of radio-compact red and blue QSOs with $0\farcs2$ resolution e-MERLIN data, we found that the radio enhancement occurs on nuclear--galactic scales~\cite{rosario_21}. Overall, these results are inconsistent with red QSO being blue QSOs observed at a higher inclination angle, whereby the line of sight intersects with the top of the torus. If~this were the case, we would expect the opposite result; i.e.,~enhanced radio emission in blue QSOs due to Doppler boosting arising from a more face-on orientation of the radio jet. Therefore, there must be factors other than orientation causing this enhancement, e.g.,~an enhancement in small-scale jets or winds causing radio-emitting shocks in the interstellar medium (ISM), enhanced star-formation (SF), or~differences in the accretion~properties.

In this paper, we undertake, for~the first time, self-consistent analyses using four different radio datasets with the same QSO parent quasar sample, to~ensure there are no differences in the construction of our samples that could be driving the radio results~\endnote{Our previous studies had small differences in the selection of the QSO samples: e.g., different SDSS data releases and different sized control samples.}. In~Section~\ref{sec:method}, we discuss the radio data and our self-consistent selection of the QSO color-selected samples used in this study. In~Section~\ref{sec:enhance}, we use these data to explore the radio enhancement in red QSOs across four orders of magnitude in radio flux over radio frequencies spanning 144\,MHz--3\,GHz, and,~in Section~\ref{sec:morph}, we compare the radio morphological sizes of red and blue QSOs. In~Section~\ref{sec:dis}, we discuss the potential origin of the radio emission.
Throughout our work, we adopt a flat $\Lambda$-cosmology with $H_0$\,$=$\,70~kms$^{-1}$Mpc$^{-1}$, $\Omega$\textsubscript{M}\,$=$\,0.3, and $\Omega_{\Lambda}$\,$=$\,0.7.

\section{Materials and~Methods}\label{sec:method}
In the following sections, we briefly discuss the radio data used in this paper \linebreak(\mbox{Section~\ref{sec:radio}}), the~optical data used to construct our QSO parent sample (\mbox{Section~\ref{sec:optical}}), and our final color-selected and radio-detected samples (Section~\ref{sec:colour}). 
\begin{figure}[H]
    \center
    \includegraphics[width=3.8in]{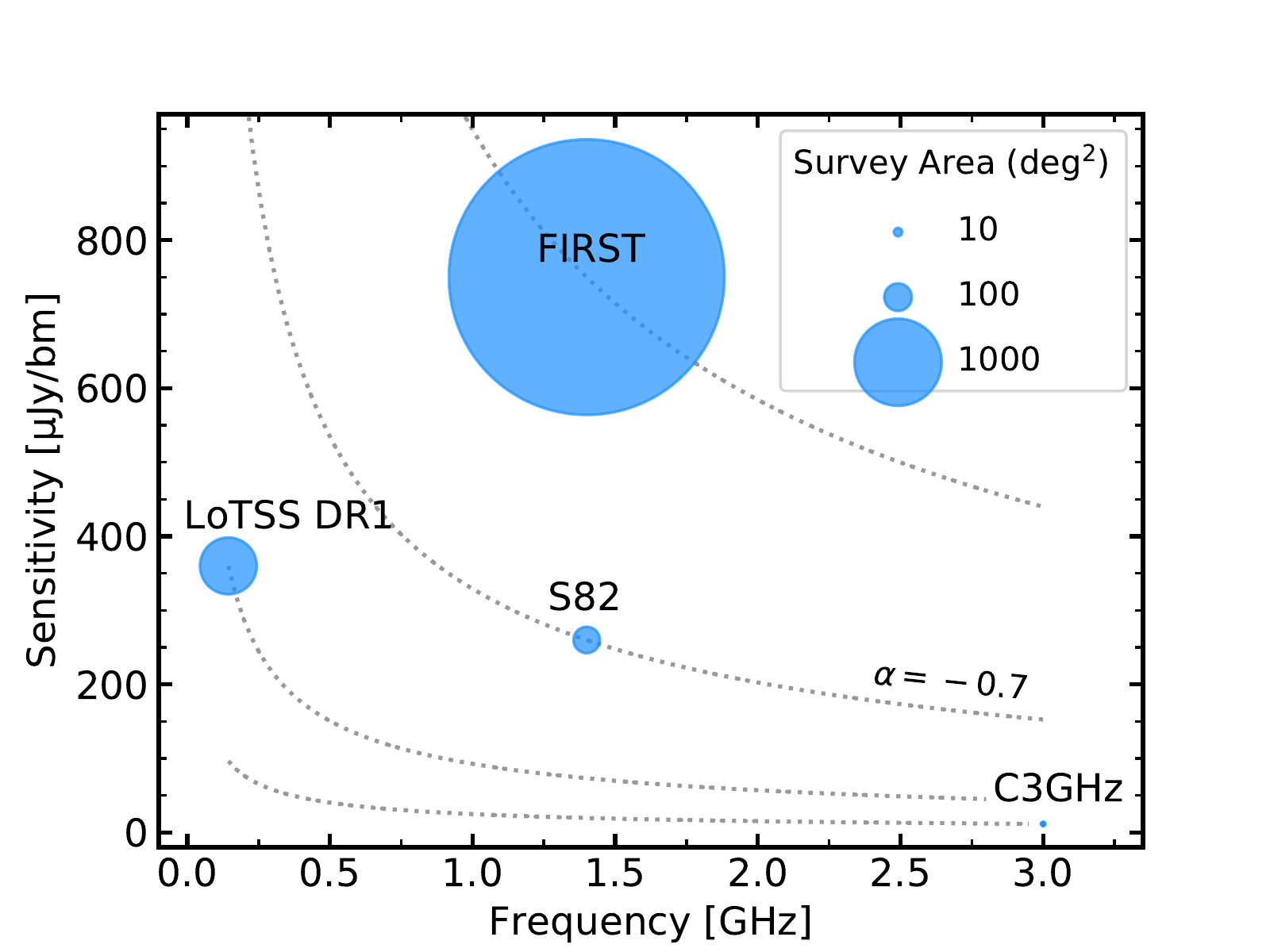}
    \caption{The $5\upsigma$ sensitivity versus radio frequency for the four radio surveys considered in this paper. The~survey area is indicated by the size of the marker. The~dotted grey lines indicate the equivalent radio sensitivities of each of the four surveys at different frequencies for compact radio sources with an $\upalpha=-0.7$: for an equivalent 1.4\,GHz luminosity, the~S82, LoTSS DR1, and C3GHz survey sensitivities are $\sim$3, 10, and 38 times deeper than that of FIRST, respectively.}
    \label{fig:context}
\end{figure}

\subsection{Radio~Data}\label{sec:radio}
An illustration of the sensitivity, frequency and sky coverage for each radio survey used in this paper is shown in Figure~\ref{fig:context}. This highlights how some radio surveys cover a large fraction of the sky, but~are lacking in sensitivity (e.g., FIRST), whereas others are extremely sensitive, but~only cover a very small area of the sky (e.g., C3GHz).

\subsubsection{VLA~FIRST}
The Faint Images of the Radio Sky (FIRST) radio survey covers 9055~deg$^2$ of the North Galactic Cap and Equatorial Strip in the SDSS region. FIRST has a 5$''$ resolution at 1.4~GHz taken primarily in the VLA B-configuration. The~catalogue contains 946,000 sources with a typical sensitivity of 0.15~mJy/bm; 30\% of the FIRST sources have optical counterparts in the SDSS \citep{first}.

\subsubsection{VLA Stripe 82 (S82)}\label{sec:s82}
The VLA Stripe 82 (S82;~\cite{hodge}) survey spans $\sim$92~deg$^{2}$ at 1.4\,GHz, centered on the equatorial SDSS Stripe 82 field (RA\,=\,20~h to 4~h, Dec\,=\,$-1.26^{\circ}$ to $1.26^{\circ}$ \cite{jiang}). S82 has a 1$\farcs$8 spatial resolution taken primarily in the A-configuration. At~its median sensitivity of 52\,$\upmu$Jy/bm, it is roughly three times deeper than~FIRST. 

In the published S82 catalogue, peak flux densities (\textit{F}\textsubscript{peak}) are derived by fitting an elliptical Gaussian model to the source. Matching to the FIRST catalogue recovers over 97\% of the FIRST-detected QSOs in this region~\cite{hodge}.

\subsubsection{VLA-COSMOS 3\,GHz (C3GHz)}\label{sec:c3ghz}
The VLA-COSMOS 3\,GHz (C3GHz) survey spans 2.6\,deg$^2$ at 3\,GHz, centered on the COSMOS field (RA\,=\,10:00:28.6, Dec\,=\,+02:12:21.0). C3GHz has a $0\farcs75$ spatial resolution taken in the A+C configuration \citep{smol}. At~its median sensitivity of $\sim$2.3\,$\upmu$Jy/bm, C3GHz is equivalently $\sim$13 times deeper than the S82 and $\sim$38 times deeper than FIRST at 1.4\,GHz, assuming a spectral index of $\upalpha$\,$=$\,$-$0.7 ($S_{\upnu}\propto \upnu^{\upalpha}$). In~the published catalogue, \textit{F}\textsubscript{peak} is measured by fitting a two-dimensional parabola around the brightest pixel. On~the basis of a Monte Carlo method, the~source completeness is 55\% up to 20\,$\upmu$Jy, which rises to 94\% above 40\,$\upmu$Jy~\cite{smol}.

The COSMOS field has also been observed at 1.4\,GHz over 2\,deg$^{2}$ in an earlier VLA radio survey \citep{smol07}. This survey has a resolution of 1$\farcs$5\,$\times$\,1$\farcs$4 and a sensitivity of $\sim$10.5\,$\upmu$Jy/bm ($\sim$5 times deeper than S82). To~calculate the $L_{\rm 1.4\,GHz}$, the~1.4\,GHz flux from this survey is used, when available, and~otherwise scaled from the 3\,GHz flux using a spectral index of $\upalpha=-0.7$.



\subsubsection{LoTSS~DR1}\label{sec:lotss}
The LOFAR Two-meter Sky Survey (LoTSS) is a 120--168 MHz survey of northern sky at a resolution of 6$''$ \citep{lotss}. The~first full-resolution data release (DR1~\cite{lotss2}) covers 424\,deg$^{2}$ in the HETDEX spring field (RA\,=\,10 h 45 m to 15 h 30 m, Dec\,=\,$45^{\circ}$ to $47^{\circ}$). At~it's median sensitivity of 71\,$\upmu$Jy/bm, LoTSS is equivalently $\sim$3.5 times deeper than S82 and $\sim$10~times deeper than FIRST at 1.4\,GHz, assuming a spectral index of $\upalpha$\,$=$\,$-$0.7.
To calculate the $L_{\rm 1.4\,GHz}$, the~144\,MHz fluxes are scaled to 1.4\,GHz using a spectral index of $\upalpha=-0.7$. 


\subsection{Optical Data: The SDSS DR14 Quasar~Catalogue}\label{sec:optical}
The SDSS DR14 Quasar Catalogue contains 526,356 spectroscopically-selected QSOs across a 9376\,deg$^{2}$ region, out to redshifts around $z=7$ \citep{paris}. The~catalogue includes previous spectroscopically-confirmed QSOs from the SDSS-I and II Legacy surveys \citep{york} with QSOs targeted by the Baryon Oscillation Spectroscopic Survey (BOSS \citep{boss}) in SDSS-III~\citep{eise} and the extended Baryon Oscillation Spectroscopic Survey (eBOSS \citep{eboss}) in~SDSS-IV. 

In this paper, the~SDSS five-band optical photometry ($u^*g^*r^*i^*z^*$) is also utilized, corrected by the associated band-dependent Galactic extinction estimates. The~quasar catalogue provides spectroscopic redshifts based on different estimators~\cite{paris}; in this work, we use the most robust of these estimates (listed as $Z$ in the catalogue).


\subsection{Color-Selected and Radio-Detected~Samples}\label{sec:colour}

\begin{figure}
    \center
    \includegraphics[width=3.5in]{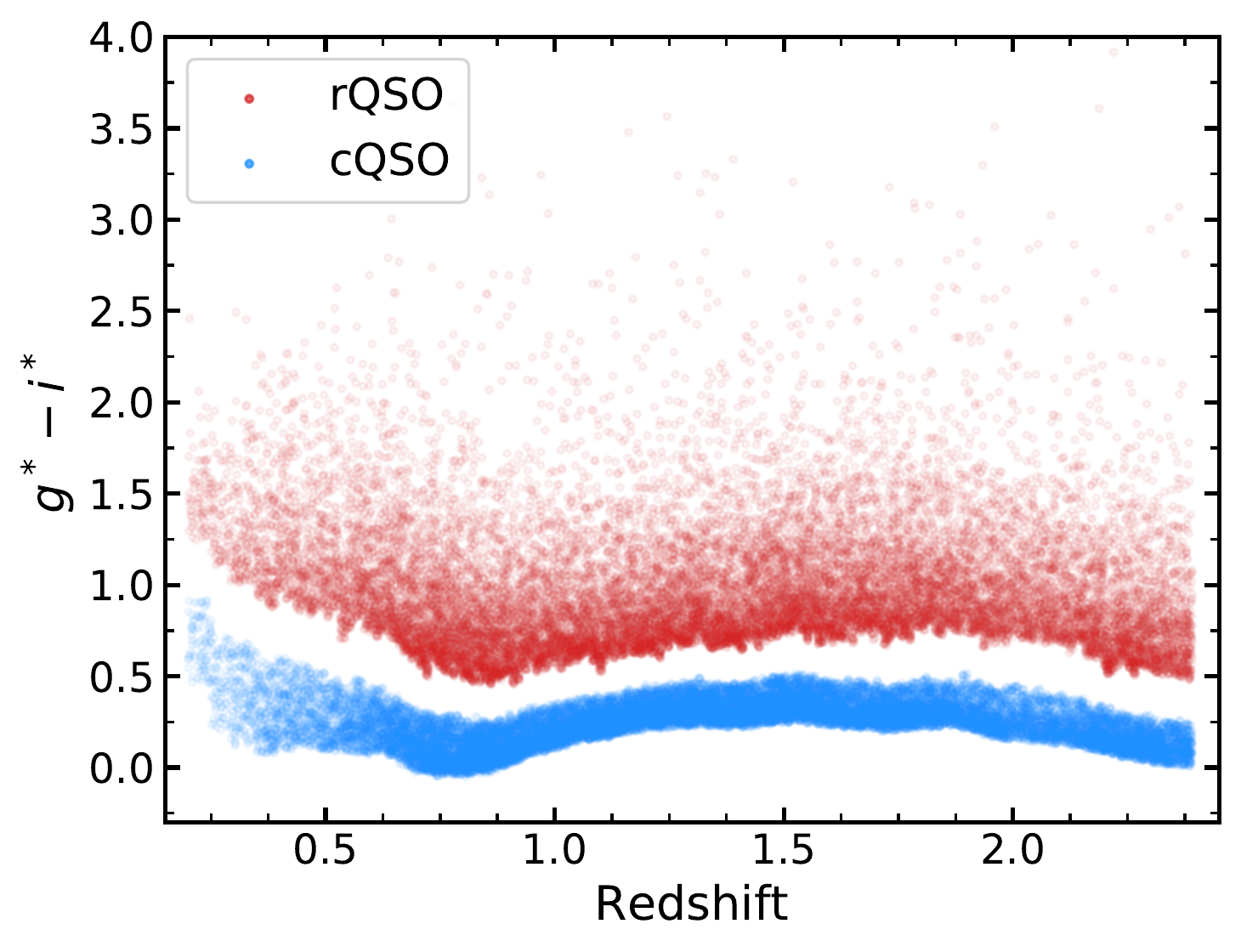}
    \caption{$g^*-i^*$ versus redshift for our parent, luminosity-redshift matched rQSO (red) and cQSO (blue) samples, which comprise the optically reddest 10\% and the middle 50\% of the $g^*-i^*$ distribution within each redshift~bin.}
    \label{fig:z_gi}
\end{figure}

To define our sample of color-selected QSOs we used the $g^*$ (4770\,\AA) and $i^*$ (7625\,\AA) band extinction-corrected photometry in the DR14 catalogue and followed a similar approach to our previous studies~\cite{klindt,fawcett,rosario}. First, we restricted to a redshift range of $0.2<z<2.4$ to ensure our color selection is not affected by the Lyman break. In~order to control for the impact of redshift on the $g^*-i^*$ color selections, we sorted the QSOs by redshift and constructed $g^*-i^*$ distributions in contiguous bins of 1000 sources. Our red and control QSOs (hereafter, rQSOs and cQSOs) were then selected above the top 90th percentile and within the middle 50th percentile of the observed SDSS $g^*-i^*$ distribution, respectively; therefore, the cQSOs represent typical QSOs in terms of their color. For~our study, here, the~color-selected QSO sample was then restricted to the various radio survey regions, using non-zero local root mean square (RMS) values either provided through a radio mosaic or an online matching~server.

Due to the loss of flux at UV-optical wavelengths, dust-reddened optically-selected rQSOs will be biased towards higher luminosities compared to blue QSOs at similar redshifts. Therefore, in~order to ensure that our results are not due to differences in the intrinsic luminosity distributions of rQSOs and cQSOs, we also matched the samples in both redshift and luminosity: for this, we use the rest-frame 6 $\upmu$m luminosity (\textit{L}\textsubscript{6$\upmu$m}) since this provides an extinction-insensitive measure of the intrinsic luminosity (e.g., Reference \mbox{\citep{stern05,lacy05,stern12,assef}}). To~calculate \textit{L}\textsubscript{6$\upmu$m}, we identified mid-infrared (MIR) counterparts from the the Wide-Field Infrared Survey Explorer (\textit{WISE} \citep{wise}), an~all-sky survey which provides photometry in four bands (3.4, 4.6, 12 and 22\,$\upmu$m). We used the NASA/IPAC query engine to match the SDSS DR14 QSOs to the All-Sky \textit{WISE} Source Catalogue (ALL-\textit{WISE}), adopting a $2\farcs7$ search radius, which ensures a 95.5\% positional certainty \citep{lake11}, and~required a detection with a signal-to-noise ratio (SNR) of greater than 2 in the \textit{WISE} \textit{W1}, \textit{W2}, and \textit{W3} bands in order to derive an accurate estimate of their \textit{L}\textsubscript{6$\upmu$m}. For~the final color-selected samples, we match in \textit{L}\textsubscript{6$\upmu$m} and redshift, using a tolerance of 0.2 and 0.05\,dex, respectively~\endnote{Due to a lack of sources, we are unable to match in luminosity for the S82 or C3GHz samples.}, following a similar approach to our original work (Section~2.3.2 in Reference~\cite{klindt}). The~$g^*-i^*$ color versus redshift distribution for our \textit{L}\textsubscript{6$\upmu$m}--$z$ matched rQSO and cQSO parent samples are shown in Figure~\ref{fig:z_gi}. The~$g^*-i^*$ distribution shows that, despite encompassing 50\% of the parent QSO population, the~cQSOs cover a narrow region of color space, compared to the broad swathe of color space encompassed by the top 10\% rQSOs.

The resulting number of color-selected and radio-detected rQSOs and cQSOs for each of the four radio surveys explored in this work are displayed in Table~\ref{tab:sample}; the radio-detection fraction and enhancement~\endnote{Defined as the radio-detection fraction of the rQSOs, divided by the radio-detection fraction of the cQSOs.} in the rQSOs compared to the cQSOs are also shown. To~quantify how many of the rQSOs and cQSOs in our sample are `radio-quiet', we adopted the same ``radio-loudness'' parameter ($\mathcal{R}$) as that first used in Reference~\cite{klindt}, defined as the dimensionless~quantity:
\begin{equation}
\mathcal{R}=\textrm{log\textsubscript{10}}\frac{\textit{L}\textsubscript{1.4\,GHz}}{\textit{L}\textsubscript{6$\upmu$m}} .\
\end{equation}

We also used the same radio-loud/radio-quiet threshold of $\mathcal{R}$\,$=$\,$-4.2$, equivalent to a mechanical-to-radiative power ratio of \textit{P}\textsubscript{mech,sync}/\textit{P}\textsubscript{rad,$L_{\rm 6\upmu m}$}\,$\approx$\,0.001, which is broadly consistent with the classical radio-quiet/radio-loud threshold (often defined using a 5\,GHz-to-2500\,\AA~flux ratio), but~is less susceptible to obscuration from dust (see Reference \citep{klindt} for full details).

 
\begin{specialtable}[H]
   \caption{The columns from left to right display the: (1) radio frequency of each dataset, (2) radio survey area, (3) number of rQSOs and cQSOs in the parent samples within each radio survey region, (4) the number of radio-detected rQSOs and cQSOs (the radio-detection fractions for each color-selected sample is displayed in the brackets), and~(5) the radio enhancement of rQSOs over cQSOs; the radio enhancement for the rQSOs is $>$1 in each sample. It should be noted that the samples S82 and C3GHz in italics are not matched in $L_{\rm 6\upmu m}$ due to lack of~sources.}
\setlength{\tabcolsep}{2.06mm}     
    \begin{tabular}{cccccccc}
        \toprule 
  
         & \boldmath{$\upnu$ }& \textbf{Area} & \multicolumn{2}{c}{\textbf{Color-Selected}} & \multicolumn{2}{c}{\textbf{Radio-Detected}} & \textbf{Radio}\\
         \textbf{Sample} & \textbf{[GHz] }& \textbf{[deg\boldmath{$^{2}$]} }& \textbf{rQSOs} & \textbf{cQSOs} & \textbf{rQSOs} & \textbf{cQSOs} & \textbf{Enhance}.\\
         \midrule
         FIRST & 1.4 & 10,000 & 20,546 & 20,546 & 2339 (11.4\%) & 940 (4.6\%) & 2.5 \\
         \textit{S82} & 1.4 & 92 & 372 & 1668 & 61 (16.4\%) & 82 (4.9\%) & 3.3 \\
         \textit{C3GHz} & 3 & 2.6 & 10 & 29 & 8 (80.0\%) & 20 (69.0\%) & 1.2 \\
         LoTSS & 0.144 & 424 & 2107 & 2107 & 761 (36.1\%) & 490 (23.3\%) & 1.6 \\
 
        \bottomrule
   \end{tabular}
   \label{tab:sample}
\end{specialtable}

\section{Results}
In this section, we use the same color-selected parent QSO sample, matched in both redshift and 6$\upmu$m luminosity, in~order to apply self-consistent analyses of the radio properties of rQSOs, using radio data from four different radio surveys, spanning radio frequencies of 0.144--3\,GHz and equivalent 1.4\,GHz radio fluxes of $\sim$0.01--1000\,mJy.

\subsection{Radio Enhancement in~rQSOs}\label{sec:enhance}
\begin{figure}[H]
\center
\includegraphics[width=4.2in]{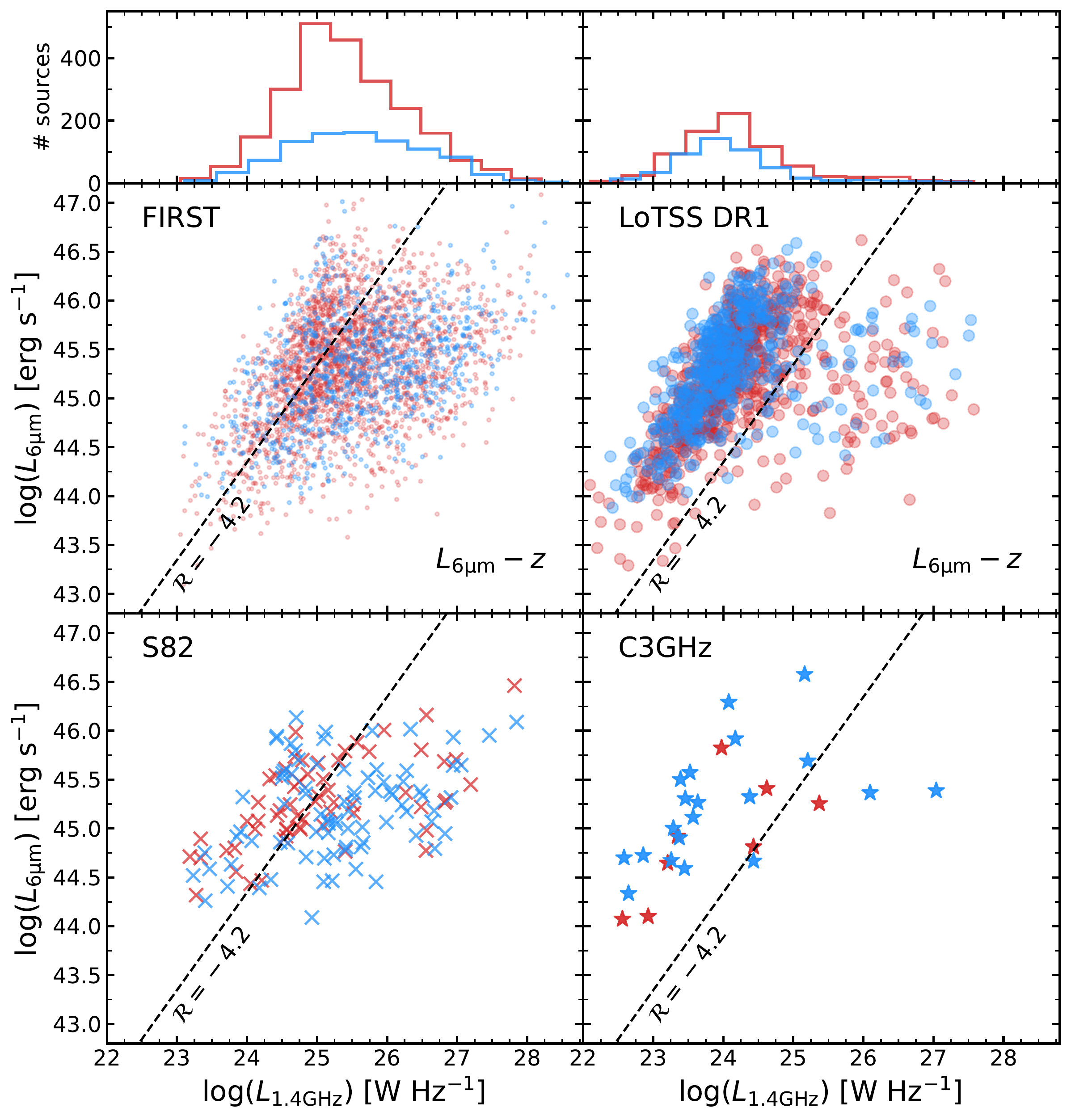}
    \caption{$L_{\rm 6\upmu m}$ versus $L_{\rm 1.4\,GHz}$ of the radio-detected FIRST (dots), LoTSS (circles), S82 (crosses), and~C3GHz (stars), rQSOs (red), and cQSOs (blue). Our division between radio-loud and radio-quiet sources ($\mathcal{R}=-4.2$) is displayed as a dashed line. To~calculate the equivalent 1.4\,GHz luminosity for the LoTSS and C3GHz samples, we assume a spectral index of $\upalpha=-0.7$ (see Figure~\ref{fig:context}). The~histograms display the 1.4\,GHz distributions for the FIRST and LoTSS samples: there is a higher number of radio-detected rQSOs compared to cQSOs. Due to the small survey area, we do not match in $L_{\rm 6\upmu m}$ for the S82 and C3GHz~samples.}
    \label{fig:radio}
\end{figure}

\begin{figure}
    \center
    \includegraphics[width=4.in]{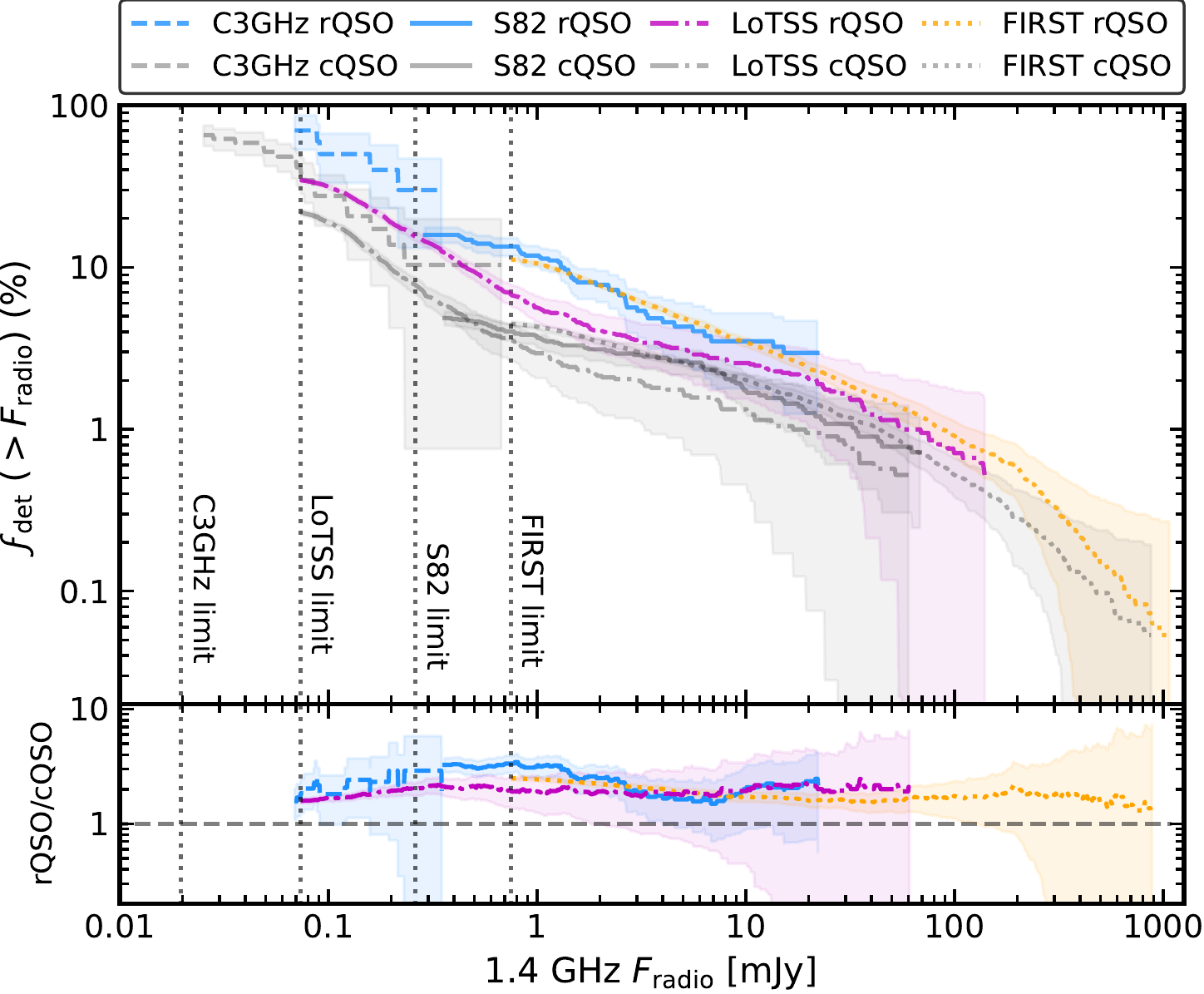}
    \caption{Cumulative radio-detection fraction as a function of 1.4\,GHz flux for the four rQSO and cQSO radio-detected samples. The~bottom panel displays the factor 2--3 times radio enhancement of the rQSOs; across 4 orders of magnitude a larger fraction of rQSOs are radio detected than compared to cQSOs. The~shaded error region was calculated using the method described in Reference~\cite{cam} and corresponds to 1$\upsigma$ binomial~uncertainties.}
    \label{fig:det_frac} 
\end{figure}

Figure~\ref{fig:radio} displays the \textit{L}\textsubscript{6$\upmu$m} versus \textit{L}\textsubscript{1.4\,GHz} distributions for the four radio-detected samples used in our analyses. The~differences in the distributions of the data is related to the depth and areal coverage of the radio surveys. The~C3GHz has the most sensitive radio data, probing down to very low $\mathcal{R}$ values, but~covers a small area and consequently few luminous radio-sources are detected. By~comparison, the~other surveys cover large areas and detect QSOs across a broad range of the $\mathcal{R}$ plane, providing good coverage overall particularly from the sensitive LoTSS survey.
Figure~\ref{fig:det_frac} displays the cumulative radio-detection fraction of our four samples as a function of the 1.4\,GHz radio flux. Across four orders of magnitude in 1.4\,GHz flux, the~rQSOs have a factor 2--3 higher radio-detection fraction in all four radio~samples.

\begin{figure}[H]
    \center
    \includegraphics[width=4.in]{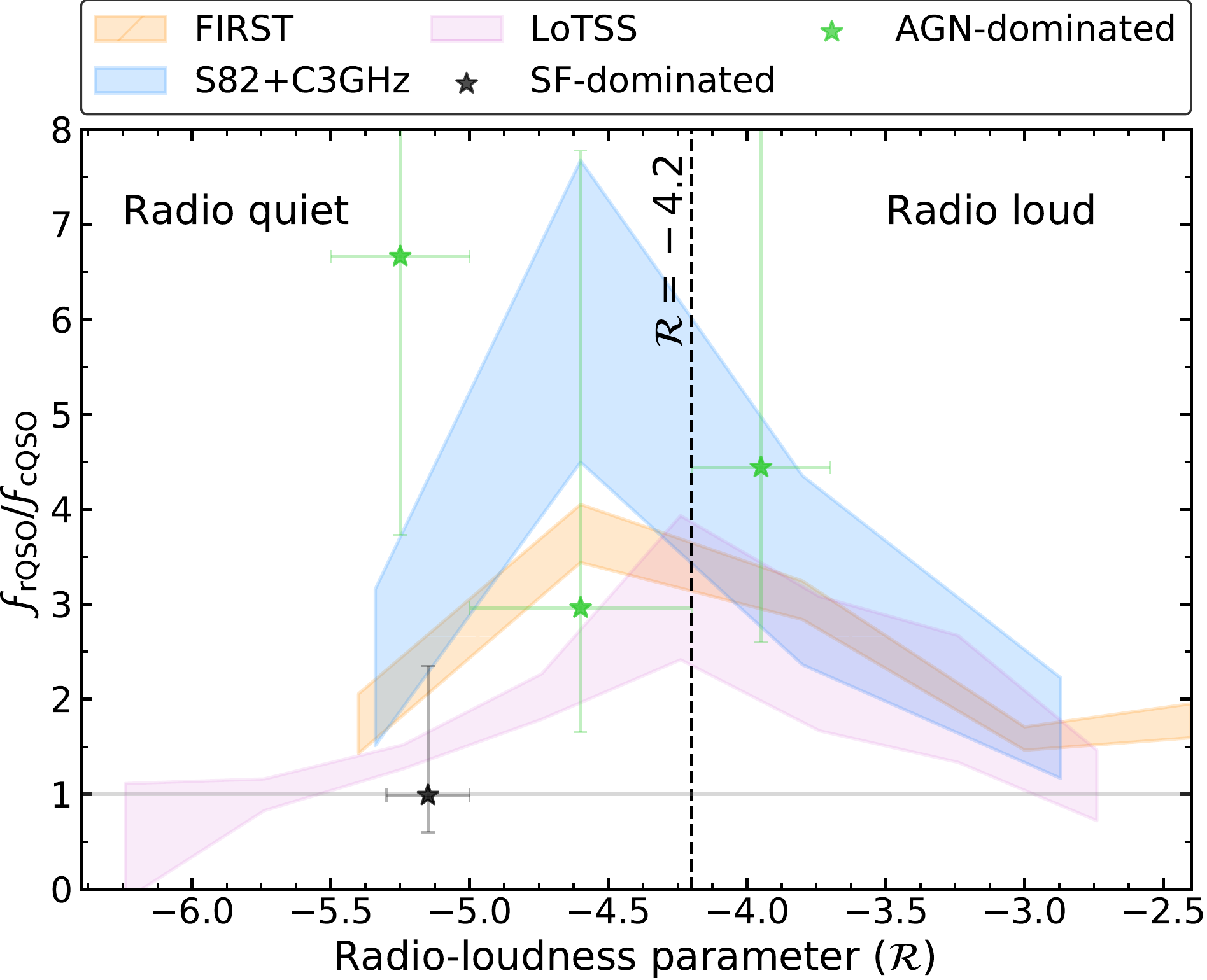}
    \caption{Radio-detection enhancement as a function of radio-loudness for the FIRST (orange), LoTSS (magenta), and~combined S82 and C3GHz (blue) samples. Our defined radio-quiet threshold ($\mathcal{R}=-4.2$) is displayed as a dashed vertical line. The~combined S82 and C3GHz data points are taken from Reference~\cite{fawcett} and the LoTSS data points are taken from Reference~\cite{rosario_21}. The~FIRST data points are calculated using our new self-consistent sample from this study. The~shaded error region was calculated using the method described in Reference~\cite{cam} and corresponds to 1$\upsigma$ binomial uncertainties: in all four samples, the~peak and decrease in the radio-detection enhancement is statistically significant. The~radio enhancement of the C3GHz rQSOs with radio emission dominated by SF is shown as a black star, and~those with radio emission dominated by AGN processes are shown as green~stars.}
    \label{fig:radio_loud}
\end{figure}

Splitting the radio-detected samples into contiguous $\mathcal{R}$ bins, we calculated the enhancement in the radio-detection fraction of the full color-selected sample for the rQSOs. This is calculated as the ratio of the radio-detection fraction of the rQSOs to the radio-detection fraction of the cQSOs in each sample (see our previous studies for more details~\cite{klindt,fawcett,rosario}). Figure~\ref{fig:radio_loud} displays the radio enhancement for the FIRST- (shaded orange), combined S82+C3GHz- (shaded blue), and LoTSS- (shaded magenta)-detected samples. All three of these samples peak around $\mathcal{R}\sim -4.5$ with a factor $\sim$3--6 excess in the radio enhancement for rQSOs, decreasing to around unity towards the extreme radio-loud and radio-quiet values; we will explore this decrease in Section~\ref{sec:SF}. The~spread in the shaded regions represent the error bars, showing both the peak and decrease in radio-detection enhancement is statistically significant for all the samples. Expanding on the work from Reference~\cite{klindt}, who find an enhancement in the FIRST-detection fraction of SDSS DR7 rQSOs around the radio-quiet/radio-loud threshold, we confirm this result at a much higher significance with our DR14 sample. We also push to lower $\mathcal{R}$ values where we find a decrease in radio-detection enhancement, consistent with what has been seen previously with deeper radio samples~\cite{fawcett,rosario}, but~now, for the first time, with the shallower FIRST~data.

\vspace{-6pt}

\subsection{Radio~Morphology}\label{sec:morph}
The majority of the radio-detected rQSOs that contribute to the radio-excess shown in Figure~\ref{fig:radio_loud} have compact morphologies at the resolution of FIRST (5$''$; $\sim$43 kpc at $z=1.5$ \cite{klindt}). However, these size constraints are just upper limits, and~higher-resolution radio data has shown that the rQSOs are typically compact over galaxy scales (e.g., $<$16\,kpc on the basis of $1\farcs8$ resolution data in the S82 field~\cite{fawcett}). 

\begin{figure}[H]
    \center
    \includegraphics[width=4.5in]{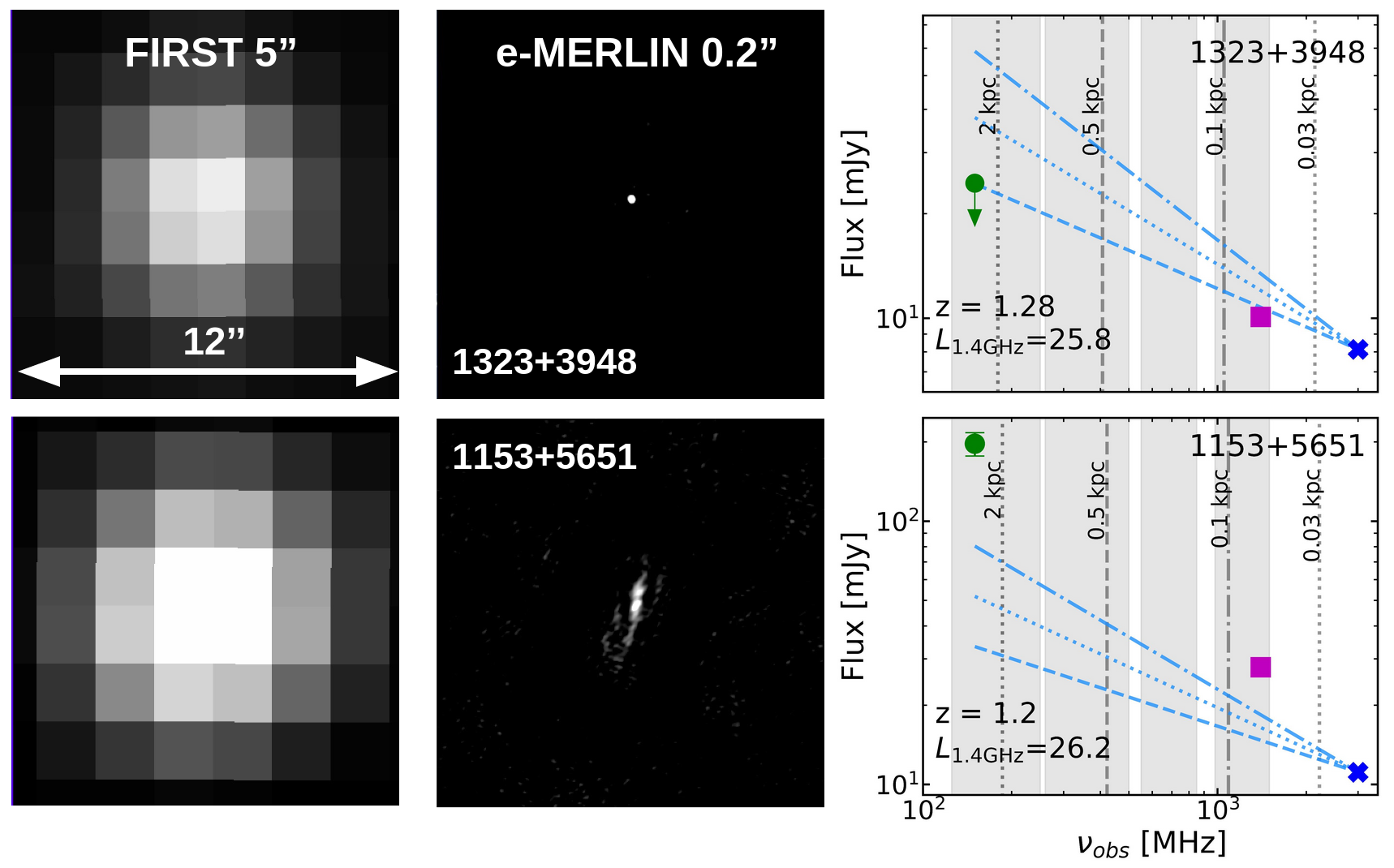}
    \caption{This figure shows $12''\times12''$ thumbnails of two rQSOs unresolved at FIRST 5$''$ resolution (left), observed in the e-MERLIN $0\farcs2$ data (middle) from Reference~\cite{rosario_21}; the top source remains unresolved at the e-MERLIN resolution but the bottom source shows $\sim$kpc-scale extended radio emission. (Right) radio flux versus frequency for the two sources displayed in the thumbnails, using 150\,MHz data from TGSS (green circle), 1.4\,GHz data from FIRST (magenta square), and 3\,GHz data from VLASS (blue cross); a 7$\upsigma$ upper limit of 24.5\,$\upmu$Jy is used for the TGSS-undetected source (top). The~grey shaded regions demonstrate the four radio bands over which we have recently obtained uGMRT data for all of our e-MERLIN observed sources, allowing us to construct sensitive radio SEDs (PI: V. Fawcett). The~dashed and dotted blue lines indicate the expected fluxes for a spectral slope of $\upalpha=-0.5$, $-0.7$, and $-0.9$, demonstrating the variety in radio spectral slopes within the FIRST-compact sources. The~vertical lines show the frequencies expected for GPS/CSS-like sources of 0.03, 0.1, 0.5, and 2\,kpc extents. This demonstrates that, for~the unresolved e-MERLIN source, with~the uGMRT data, we will be able to constrain the radio sizes to $<$2 kpc scales if it is a GPS/CSS-like source, below~the resolution of the e-MERLIN~data.}
    \label{fig:size}
\end{figure}
\begin{figure}[H]
    \center
    \includegraphics[width=3.5in]{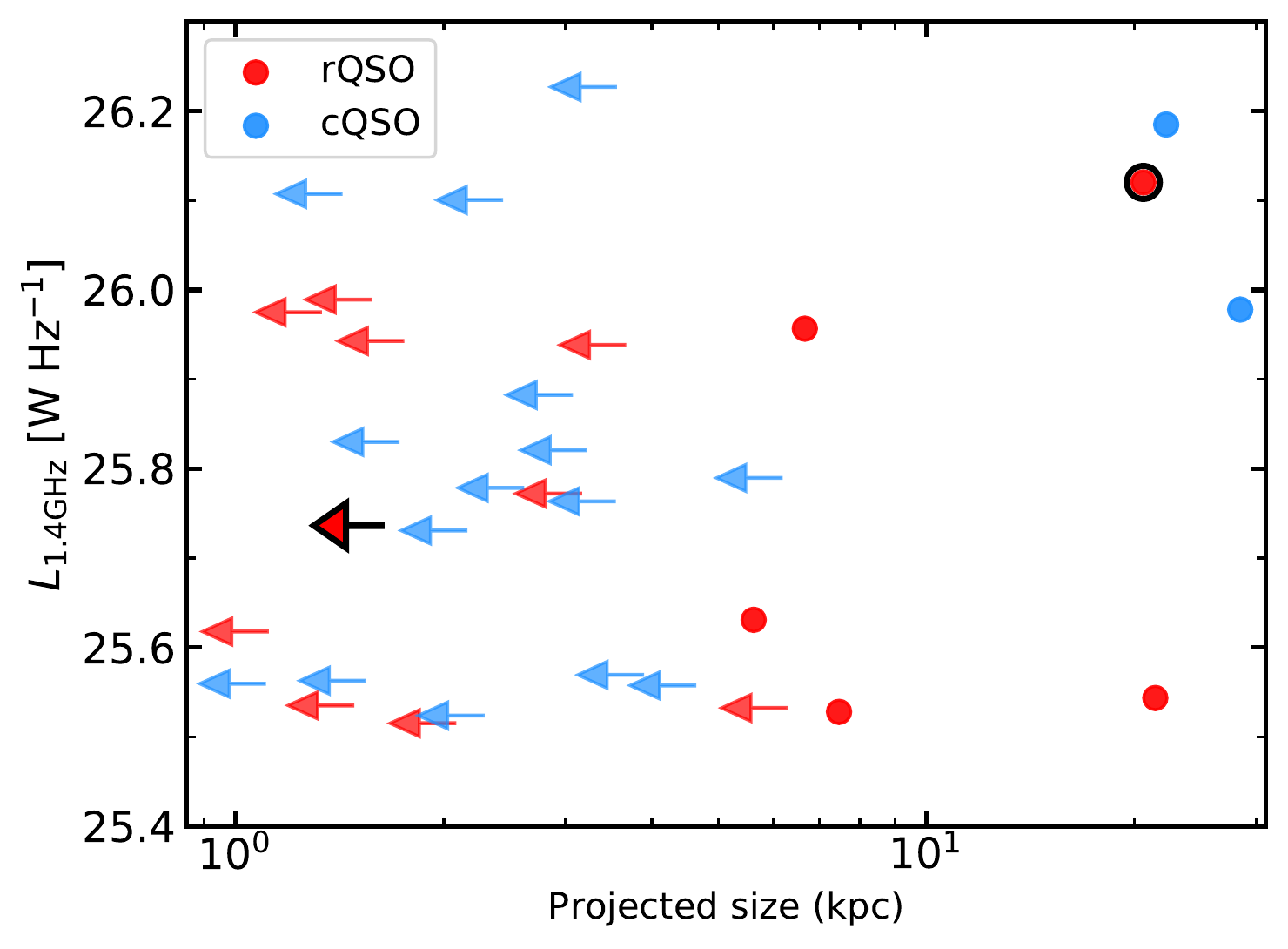}
    \caption{$L_{\rm 1.4\,GHz}$ versus projected size of the radio emission for the e-MERLIN rQSOs and cQSOs from Reference~\cite{rosario_21}. The~resolved sources are displayed as circles and the unresolved sources are shown as upper limits. The~two rQSOs displayed in Figure~\ref{fig:size} are outlined in black: source 1153+5651 is resolved, and 1323+3948 is~unresolved.}
    \label{fig:radio_lum}
\end{figure}

Most recently, we have obtained $0\farcs2$ e-MERLIN radio data for 19 and 20 FIRST radio-compact rQSOs and cQSOs, finding that the majority remain unresolved (suggesting $<$2\,kpc scales) but also revealing a statistically significant enhancement in the fraction of rQSOs that are resolved over the $\sim$2--10\,kpc scales of the host galaxy~\cite{rosario_21}. A~comparison of the radio image of two rQSOs at FIRST 5$''$ resolution and e-MERLIN $0\farcs2$ resolution is shown in the left of Figure~\ref{fig:size}: in the lower example, there is clear extended emission at e-MERLIN resolutions, while the upper example remains unresolved. Figure~\ref{fig:radio_lum} displays the $L_{\rm 1.4\,GHz}$ versus projected size of the e-MERLIN rQSOs and cQSOs; resolved sources are shown as circles, and unresolved sources are shown as upper limits. The~resolved and unresolved source displayed in the thumbnails of Figure~\ref{fig:size} are outlined in black. This demonstrates that radio data sensitive to smaller spatial scales (e.g., VLBA observations at $\sim$0$\farcs$01) is required to directly measure the resolved sizes for the majority of the sources that contribute to the radio excess (see Section~\ref{sec:agn} for a more detailed discussion).

\unskip

\section{Discussion}\label{sec:dis}
In this paper, we have reviewed our recent research that identified fundamental differences in the radio properties between rQSOs and cQSOs. We also undertook a series of new analyses using the same color-selected parent QSO sample, which we then matched to four different radio surveys at various depths and frequencies. This allowed us to perform self-consistent analyses based on the same QSO sample across the different radio datasets. We confirmed our previous radio results, finding an enhancement in the radio-detection fraction of rQSOs compared to cQSOs that arises in radio compact systems around the radio-quiet threshold \citep{klindt,fawcett,rosario,rosario_21}.
This small-scale enhanced radio emission found in red QSOs could be driven by SF, AGN processes, or differences in the accretion properties. In~a recent study (Fawcett et~al. (\textit{submitted})), we used broad-wavelength Very Large Telescope (VLT) X-shooter~\cite{vernet} spectra to explore the accretion properties of rQSOs and cQSOs, finding no significant differences between the rQSOs and cQSOs. Consequently, it appears unlikely that the radio differences between the rQSOs and cQSOs in our SDSS studies are driven by differences in the accretion properties. Therefore, in~the following sections, we restrict our discussion to SF and AGN processes as drivers of the enhanced radio emission in~rQSOs.

\subsection{Origin of the Radio Emission: Star-Formation}\label{sec:SF}
It has been shown in previous studies that SF dominates the radio emission in the \textls[-11]{majority of classically radio-quiet quasars, although~many can still be \mbox{AGN/jet-powered~\cite{condon,herrera,macfarlane}}}. In~our previous study, we analyzed the spectral energy distributions (SEDs) for a sample of rQSOs and cQSOs and found no differences in the average star-formation rate properties~\cite{calistro}; this will be robustly tested in our future ALMA study. Therefore, it is not obvious that SF should be driving the differences we see in the radio properties of red and blue QSOs. In~order to further test this, in~an earlier study, we utilized the rest-frame (8--1000\,$\upmu$m) far-infrared (FIR) data available in the COSMOS field to provide constraints on SF~\cite{fawcett}. We defined the origin of the radio emission as dominated by SF if the measured radio luminosity was within a factor of 3 of the radio-FIR relationship for star-forming galaxies, and AGN-dominated otherwise. Using these definitions, we then split the lower radio-loudness bins covered by our C3GHz sample into sources with radio emission either dominated by SF or AGN (black and green stars, respectively; see Figure~\ref{fig:radio_loud}). From~Figure~\ref{fig:radio_loud}, it is clear that, at $\mathcal{R}<-5$, SF-dominated sources have a radio enhancement consistent with unity (albeit low source statistics), expected if there are no differences in the SF properties of rQSOs and cQSOs. We also found the AGN-dominated sources, although~very uncertain, are consistent with being enhanced at all values of $\mathcal{R}$, suggesting that the decrease in the radio enhancement of rQSOs at $\mathcal{R}<-5$ is due to the increased contribution of SF to the radio emission. Therefore, the~radio enhancement in rQSOs is likely driven by AGN~mechanisms.

\subsection{Origin of the Radio Emission: AGN}\label{sec:agn}
From the e-MERLIN observations of the FIRST radio-compact QSOs, we found a statistically significant excess in the fraction of rQSOs with kpc-scale extended emission when compared to the cQSOs. However, the~majority of both the rQSO and cQSO sources remained unresolved on $<$2 kpc scales. The~mechanism driving this compact radio emission could be wind-driven shocks, small-scale synchrotron jets, or larger jets with synchrotron self-absorption. A~potential signature of the latter scenario is a turnover in the radio SED due to self-absorbed synchrotron radiation from a frustrated jet, similar to Compact Steep Spectrum (CSS) and Gigahertz-Peaked Spectrum (GPS) sources, which have well-defined peaks around 100\,MHz and 1\,GHz, respectively~\cite{odea}, or~a flat spectrum due to a self-absorbed core~\cite{cotton}. Our future work analyzing 4-band 150\,MHz--1.4\,GHz upgraded Giant Meter Radio Telescope (uGMRT) data of our e-MERLIN sample (PI: V. Fawcett), in~addition to the VLA Sky Survey (VLASS~\cite{vlass}) 3\,GHz data, will be crucial for constructing sensitive 0.144--3\,GHz radio SEDs to test the main mechanism driving the radio emission. We already have some insight into what we might expect to find with the uGMRT data from our previous study, where we constrained the radio spectral slopes of rQSOs and cQSOs that were compact in LoTSS at two frequencies: 144\,MHz and 1.4\,GHz from LoTSS DR1 and FIRST, respectively~\cite{rosario}. From~this analysis, we found a broad range in spectral indices that peaked around $\upalpha\sim-0.7$, consistent with synchrotron emission by normal jets or winds, but~found no significant differences between the rQSOs and cQSOs. The~right plot of \mbox{Figure~\ref{fig:size}} displays the radio fluxes from the 150\,MHz TIFR GMRT Sky Survey (\mbox{TGSS; \citep{tgss}}), or an upper limit for the undetected unresolved source, FIRST, and~VLASS for the two sources shown in the left thumbnails. The~dashed and dotted blue lines indicate the expected fluxes for a spectral slope of $\upalpha=-0.5$, $-0.7$, and $-0.9$, and~the grey shaded regions are the radio frequencies covered by our uGMRT data, where we could see a spectral turnover in the radio SEDs of our rQSOs and~cQSOs.

In addition to providing an insight into the emission mechanism, the~turnover frequency of a peak in the radio SED also varies inversely with the linear size of the radio emission for GPS/CSS-like sources, providing sub-kpc constraints on the scale of the radio emission without the need for high resolution imaging~\cite{gps}. In~Figure~\ref{fig:size}, the vertical lines show the frequency turnovers expected for GPS/CSS-like sources of 0.03, 0.1, 0.5, and 2\,kpc extents, demonstrating that, with our uGMRT data, we will directly constrain the number of rQSOs and cQSOs that have frustrated jets on $\sim$0.1--2\,kpc scales. If~the turnover frequency occurs at $<$150\,MHz, then, this would imply synchrotron self-absorption by a frustrated jet on scales $>$2 kpc, resolvable for the majority of our sources with e-MERLIN (e.g., the~bottom panel of Figure~\ref{fig:size}). Therefore, we will be able to rule out this scenario for the unresolved e-MERLIN sources that display no spectral turnover in their radio SEDs.

\begin{figure}[H]
    \center
    \includegraphics[width=4.5in]{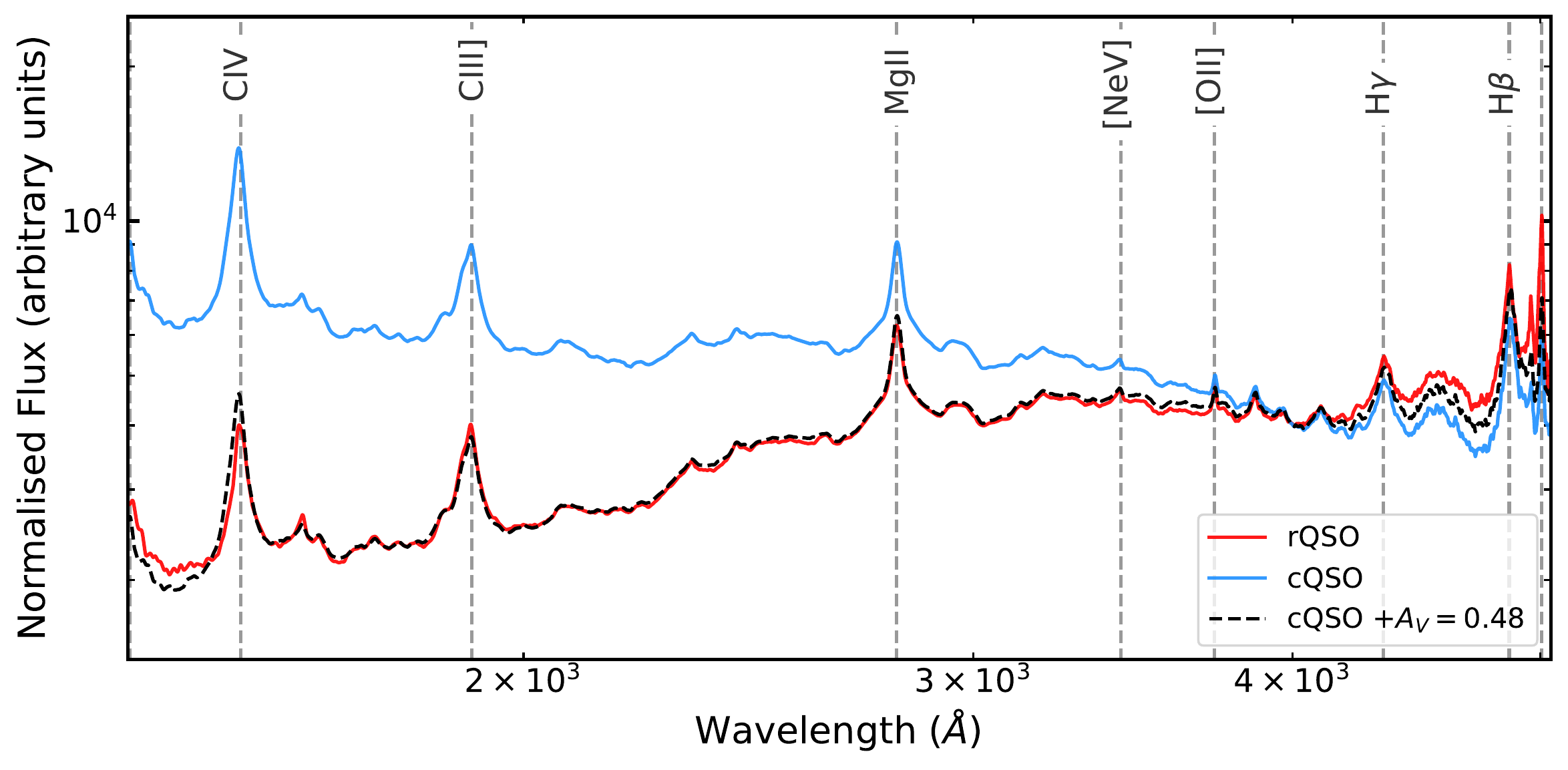}
    \caption{SDSS DR14 composites for our rQSO (red) and cQSO (blue) samples, in~a redshift range $1.0<z<1.6$ and luminosity range $45.3<\rm log$\,$L_{\rm 6\upmu m}<47.0$\,erg\,s$^{-1}$. The~dotted black line shows the cQSO composite with a dust extinction of $A_V=0.48$\,mags that is remarkably similar to the rQSO composite. This amount of dust extinction is consistent with that found from fitting the individual VLT/X-shooter spectra of a sample of rQSOs and cQSOs at $z\sim1.5$ (Fawcett~et~al. (\textit{submitted})) and in our SED analysis of SDSS rQSOs and cQSOs~\cite{calistro}. The~major emission lines are indicated by a dashed vertical grey~line.}
    \label{fig:composite}
\end{figure}

For the sources that show no evidence in their radio SED for synchrotron self-absorption, the~radio emission could be driven by either small-scale synchrotron jets or wind-driven shocks. If~the enhanced radio emission is due to small-scale synchrotron jets, then, it is not immediately clear why there would be a close connection between the radio emission and optical color. Exploring the origin of the red colors, in~Fawcett et~al. (\textit{submitted}), we use VLT/X-shooter spectra to explore the cause of the red colors and find that the reddening in rQSOs is fully consistent with moderate amounts of dust ($A_V\sim0.1$--1\,mags), as~also inferred in our recent work~\cite{calistro}. To~further demonstrate this result, Figure~\ref{fig:composite} displays DR14 SDSS rQSO (red) and cQSO (blue) median composites, in~the redshift range $1.0<z<1.6$ and luminosity range $45.3<\rm log$\,$L_{\rm 6\upmu m}<47.0$\,erg\,s$^{-1}$. A~dust-reddened cQSO composite ($A_V\sim0.5$\,mags) is shown by the dotted black line, which describes the shape of the red composite remarkably well. A~direct connection between the presence of dust and radio emission could be through shocks, whereby a dusty wind from the nucleus interacts with the ISM. We have previously found evidence for larger FWHMs in the broader wing component of the [{\sc O iii}]$\uplambda5007$ in rQSOs compared to cQSOs at $z<0.8$; evidence for stronger winds within the rQSO population~\cite{calistro}. In~Alexander et~al. (\textit{in prep}), we also find a link between rQSOs and Low-ionization Broad Absorption Line QSOs (LoBALs; BALQSOs that display additional absorption in low-ionization species, such as Mg\,{\sc ii} and Al\,{\sc iii}, and are known to host powerful outflows, e.g., Reference~\cite{trump}); LoBALs tend to have redder optical colors on average and have enhanced radio emission (also identified in other studies, e.g., Reference~\cite{balqso}). Therefore, although~we cannot rule out small-scale synchrotron jets, wind-driven shocks interacting with the ISM can more directly associate the excess radio with the presence of dust and is consistent with our previous~results.

\section{Conclusions}
We have undertaken new, self-consistent analyses using the same DR14 QSO parent sample, exploring the radio properties of a sample of $0.2<z<2.4$ red and typical QSOs. We confirm previous results but for the first time utilizing four different radio datasets at various frequencies and depths: FIRST, S82, C3GHz, and LoTSS DR1. Overall, we find:

\begin{itemize}  
    \item \textbf{Red QSOs have enhanced compact 0.144--3\,GHz radio emission that peaks around the radio-quiet threshold (see Figures~\ref{fig:det_frac}--\ref{fig:radio_lum}):} We find an enhancement in the radio-detection fraction of red QSOs across a 1.4\,GHz radio flux range of 0.1--1000\,mJy. Exploring the radio-loudness parameter, we show that the radio enhancement in all four samples, covering radio frequencies of 0.144--3\,GHz, peaks around the radio-quiet threshold and decreases towards both the extreme radio-loud and radio-quiet ends. We confirm the results from our previous work exploring the FIRST properties of SDSS DR7 red QSOs, but~now at a higher significance using a 10 times larger sample. We also push to lower radio-quiet values than previously explored, confirming the decrease in radio-detection enhancement found with the deeper radio data. We confirm that the majority of radio-compact red QSOs have radio emission on $<$2 kpc scales~\cite{rosario_21} and show that our future uGMRT study could provide sub-kpc constraints on the scale of the radio emission.
    \item \textbf{The enhanced radio emission in red QSOs is likely due to dusty winds or frustrated jets (see Section~\ref{sec:dis} and Figures~\ref{fig:size} and \ref{fig:composite}):} Comparing a red QSO SDSS composite to a moderately dust-reddened blue composite, we show that dust is the likely cause of the red colors. This result is also confirmed in our upcoming X-shooter study where we also do not find any significant differences in the accretion properties between red and blue QSOs (Fawcett~et~al. (\textit{submitted})). In~a previous study, we used the FIR data in the COSMOS field to show that the enhanced radio emission in red QSOs is likely driven by AGN processes (see Figure~\ref{fig:radio_loud}), consistent with Reference~\cite{calistro}. A~self-consistent scenario that explains these results, in~addition to the radio enhancement, is that red QSOs reside in a more dust and gas rich environment, in which the radio emission is due to wind-driven shocks or frustrated jets interacting with the ISM/circumnuclear~environment.
\end{itemize}

Understanding the connection between red QSOs and the compact radio emission promises to provide a key insight on the relationship between red and blue QSOs: are red QSOs a transition stage between obscured and unobscured QSOs, or~is the relationship more complex? We have found enhanced nuclear--galaxy scale radio emission in red QSOs compared to typical blue QSOs that could be driven by either frustrated jets, winds, or small-scale synchrotron jets. Due to the link with opacity/dust, emerging evidence from our studies suggests that the most plausible scenario is either dusty winds or frustrated jets interacting with the ISM/circumnuclear environment. With~our future uGMRT+VLASS study, we aim to robustly distinguish between these two scenarios through the construction of sensitive 0.15--3\,GHz radio~SEDs.

\vspace{6pt} 




\authorcontributions{V.A.F., D.M.A., D.J.R. and L.K. contributed equally to this manuscript subject to the following sections: data curation, V.A.F.; original draft preparation, V.A.F.; supervision, D.M.A. and D.J.R.; project administration, D.M.A. and D.J.R. All authors have read and agreed to the published version of the~manuscript.}

\funding{We acknowledge a quota studentship through grant code ST/S505365/1 funded by the Science and Technology Facility Council (VAF), the~Faculty of Science Durham Doctoral Scholarship (LK), the~Science and Technology Facilities Council (DMA, DJR, through grant codes ST/P000541/1 and ST/T000244/1).
Funding for the Sloan Digital Sky Survey has been provided by the Alfred P. Sloan Foundation, the~U.S. Department of Energy Office of Science, and~the Participating Institutions. SDSS-IV acknowledges support and resources from the Center for High-Performance Computing at the University of~Utah. }

\institutionalreview{Not applicable.} 

\informedconsent{Not applicable.}


\dataavailability{Publicly available datasets were analyzed in this study. This data can be found here: 
DR14 Quasar Catalog: \url{https://www.sdss.org/dr14/algorithms/qso\_catalog/} (accessed on 18 November 2021); 
ALLWISE Catalog: \url{https://irsa.ipac.caltech.edu/cgi-bin/Gator/nph-dd} (accessed on 18 November 2021); 
FIRST Catalog: \url{http://sundog.stsci.edu/first/catalogs.html} (accessed on 18 November 2021); 
VLA Stripe 82 Catalog: \url{http://www.physics.drexel.edu/~gtr/vla/stripe82} (accessed on 18 November 2021); 
VLA COSMOS 1.4 \& 3\,GHz Catalogs: \url{https://cosmos.astro.caltech.edu/page/radio} (accessed on 18 November 2021); 
LoTSS DR1 Catalog: \url{https://lofar-surveys.org/releases.html} (accessed on 18 November 2021); 
VLASS Catalog: \url{https://cirada.ca/catalogues} (accessed on 18 November 2021); 
TGSS Catalog: \url{http://tgssadr.strw.leidenuniv.nl/doku.php} (accessed on 18 November 2021); 
The e-MERLIN data presented in this study are available on request from the corresponding author.} 

\acknowledgments{SDSS-IV, the~primary data set used in this analysis, is managed by the Astrophysical Research Consortium for the Participating Institutions of the SDSS Collaboration including the Brazilian Participation Group, the~Carnegie Institution for Science, Carnegie Mellon University, the~Chilean Participation Group, the~French Participation Group, Harvard-Smithsonian Center for Astrophysics, Instituto de Astrof\'isica de Canarias, The Johns Hopkins University, Kavli Institute for the Physics and Mathematics of the Universe (IPMU)/University of Tokyo, the~Korean Participation Group, Lawrence Berkeley National Laboratory, Leibniz Institut f\"ur Astrophysik Potsdam (AIP), Max-Planck-Institut f\"ur Astronomie (MPIA Heidelberg), Max-Planck-Institut f\"ur Astrophysik (MPA Garching), Max-Planck-Institut f\"ur Extraterrestrische Physik (MPE), National Astronomical Observatories of China, New Mexico State University, New York University, University of Notre Dame, Observat\'ario Nacional/MCTI, The Ohio State University, Pennsylvania State University, Shanghai Astronomical Observatory, United Kingdom Participation Group, Universidad Nacional Aut\'onoma de M\'exico, University of Arizona, University of Colorado Boulder, University of Oxford, University of Portsmouth, University of Utah, University of Virginia, University of Washington, University of Wisconsin, Vanderbilt University, and~Yale~University.
This publication makes use of data products from the Wide-field Infrared Survey Explorer, which is a joint project of the University of California, Los Angeles, and~the Jet Propulsion Laboratory/California Institute of Technology, funded by the National Aeronautics and Space~Administration.
The National Radio Astronomy Observatory is a facility of the National Science Foundation operated under cooperative agreement by Associated Universities, Inc.}

\conflictsofinterest{The authors declare no conflict of interest. The~funders had no role in the design of the study; in the collection, analyses, or~interpretation of data; in the writing of the manuscript, or~in the decision to publish the~results.} 


\abbreviations{Abbreviations}{The following abbreviations are used in this manuscript:\\

\noindent 
\begin{tabular}{@{}ll}
AGN & Active Galactic Nuclei \\
BALQSO & Broad Absorption Line Quasar\\
COSMOS & The Cosmic Evolution Survey \\
cQSO & Control Quasar \\
CSS & Compact Steep Spectrum\\
C3GHz & COSMOS 3\,GHz \\
e-MERLIN & Extended Multi-Element Radio-Linked Interferometer Network \\
FIR & Far-Infrared \\
FIRST & Faint Images of the Radio Sky at Twenty-cm \\
GPS & Gigahertz-Peaked Spectrum\\
LoBAL & Low-ionization Broad Absorption Line quasar\\
LOFAR & LOw Frequency ARray \\
LoTSS & The LOFAR Two-meter Sky Survey \\
MIR & Mid-Infrared \\
QSO & Quasi-stellar object\\
rQSO & Red Quasar\\
RMS & Root Mean Square \\
SDSS & Sloan Digital Sky Survey \\
SED & Spectral Energy Distribution \\
SF & Star-formation \\
S82 & Stripe 82 \\
TGSS & TIFR GMRT Sky Survey \\
uGMRT & Upgraded Giant Meterwave Radio Telescope \\
VLA & Very Large Array \\
VLASS & VLA Sky Survey \\
VLT & Very Large Telescope \\
WISE & Wide-Field Infrared Survey Explorer \\
\end{tabular}}

 \begin{adjustwidth}{-5.0cm}{0cm}
 \printendnotes[custom]
 
 \end{adjustwidth}

\end{paracol}
\reftitle{References}

\end{document}